\documentclass[aps]{revtex4}
\usepackage{amsthm,amsmath,amssymb,amscd,epsfig,float}
\def\abs#1{\left\vert #1 \right\vert}

\def\v#1{{\mathbf #1}}
\def\u#1{\,\mathrm{#1}}

\def\trace#1{{\mathrm{Tr}\left[#1\right]}}

\def\RDM#1#2{{}^{#1}\!{#2}}

\def\refsec#1{Sec.\ \ref{Section::#1}}

\def\refeqn#1{Eq.\ (\ref{Equation::#1})}

\def\refeqs#1#2{Eqs.\ (\ref{Equation::#1}) and (\ref{Equation::#2})}
\def\refeqto#1#2{Eqs.\ (\ref{Equation::#1}--\ref{Equation::#2})}

\begin{document}
\title{The tensor hypercontracted parametric reduced density matrix algorithm: coupled-cluster accuracy with O(r$^4$) scaling}
\author{Neil Shenvi$^1$}
\author{Helen Van Aggelen$^{1,2}$}
\author{Yang Yang$^{1}$}
\author{Weitao Yang$^1$}
\author{Christine Schwerdtfeger$^{3}$}
\author{David Mazziotti$^4$}
\affiliation{$^1$ Dept. of Chemistry, Duke University, Durham, NC 27708}
\affiliation{$^2$ Dept. of Inorganic and Physical Chemistry, Ghent University, 9000 Ghent, Belgium}
\affiliation{$^3$ Dept. of Chemistry, The University of Illinois at Urbana-Champaign, Urbana, IL 61801}
\affiliation{$^4$ Dept. of Chemistry, The University of Chicago, Chicago, IL 60637}
\date{\today}

\begin{abstract}
Tensor hypercontraction is a method that allows the representation of a high-rank tensor as a product of lower-rank tensors.  
In this paper, we show how tensor hypercontraction can be applied to both the electron repulsion integral (ERI) tensor and 
the two-particle excitation amplitudes used in the parametric reduced density matrix (pRDM) algorithm.  Because 
only $O(r)$ auxiliary functions are needed in both of these approximations, our overall algorithm can be shown to 
scale as O(r$^4$), where $r$ is the number of single-particle basis functions.  We apply our algorithm to several 
small molecules, hydrogen chains, and alkanes to demonstrate its low formal scaling and practical utility.  
Provided we use enough auxiliary functions, we obtain accuracy similar to that of the traditional pRDM algorithm, 
somewhere between that of CCSD and CCSD(T).  
\end{abstract}

\maketitle

\section{Introduction} \label{Section::Introduction}

The problem of the rapid growth of the electronic wavefunction with system size has plagued quantum chemistry for 
decades\cite{Levine:90,Szabo:96}.  There have been many attempts to conquer the `curse of dimensionality' and a 
large number of highly successful approximations have been developed.  The simplest approach is the one taken by 
Hartree-Fock theory, which approximates the exact wavefunction as a single Slater determinant.  Missing from the 
Hartree-Fock result is the effect of electronic correlation, which determines many of the atomic and molecular 
properties in which chemists are most interested.  

The difficulty inherent in electronic structure calculations is that correlated methods usually scale as higher 
orders of the number of basis functions involved in the calculation.  A straightforward implementation of 
Hartree-Fock scales as $O(r^4)$ where $r$ is the number of basis functions.  Approximate methods employed to 
capture correlation face a trade-off between accuracy and efficiency\cite{Bartlett:ARPC81,HeadGordon:JPC96}.  
Perturbative treatments such as MP2 and MP4 scale as $O(r^5)$ and $O(r^6)$, 
respectively\cite{Bartlett:ARPC81}.  Configuration interaction methods, which add in double-, triple- and quadruple-order
excitations, form a hierarchy of methods which scale as $O(r^6)$ and higher\cite{Bartlett:RMP07}.  A plethora of other, highly 
accurate methods based on coupled-cluster theory\cite{Bartlett:JPC89,Bartlett:RMP07}, 2-particle reduced density 
matrices\cite{Mazziotti:JCP10,Mazziotti:PRL11,Mazziotti:JCP11,Mazziotti:CR12}, and reduced active 
space diagonalization\cite{Roos:CPL89} also scale as at least $O(r^6)$.  Due to this high $O(r^6)$ scaling, these methods 
are generally 
not applicable beyond small molecules.  Instead, quantum chemists have increasingly turned to density functional 
theory (DFT), which offers low $O(r^4)$ or even $O(r^3)$ formal scaling while managing to capture varying amounts of the electronic 
correlation\cite{Yang:89, Yang:CR12}.  Nonetheless, the search for efficient wavefunction-based methods that are competitive 
with DFT has continued, motivated by the importance of strong correlation in many systems such as transition metal clusters, 
solid state devices, and molecules far from their equilibrium geometries.

One approach which attempts to reduce both the cost and scaling of correlated electronic structure methods is the 
decomposition of the electronic repulsion integral (ERI) tensor, which is naturally a rank-4 object, into products of 
lower-rank objects.  Resolution-of-the-identity techniques\cite{Feyereisen:CPL93,Ahlrichs:CPL95,Fruchtl:TCA97,Haser:TCA97} are 
one subset of this approach, as are pseudospectral methods\cite{Martinez:JCP93,Martinez:JCP94,Martinez:JCP95}.  Recently, 
Hohenstein et al\cite{Martinez:JCP12a} introduced tensor hypercontraction density fitting (THC-DF) to decompose the electron 
repulsion integral tensor, an object which is central to {\it ab initio} electronic structure calculations.  The authors 
showed that an auxiliary basis could be used to express the rank-4 ERI tensor as the product of five rank-2 tensors.  This 
approach led to O(r$^4$) algorithms for MP2, MP3\cite{Martinez:JCP12a}, CC2\cite{Martinez:JCP12c}, and -tentatively- for CISD 
and CCSD\cite{Martinez:JCP13}, where $r$ is the number of one-particle basis functions.  In the last case of the CISD and CCSD 
algorithms, the THC methodology has not yet been implemented in a fully $O(r^4)$ manner, but initial results were 
encouraging.  In all of these cases, the application of tensor hypercontraction adds a non-negligible overhead and prefactor 
to the algorithm.  However, the reduced asymptotic scaling will lead to a cross-over that renders the THC versions more 
efficient for sufficiently large systems. 

Concurrently, Mazziotti et al. have been developing a parametric 
reduced density matrix (pRDM) algorithm that can be used to obtain ground state electronic energies with an accuracy somewhere 
between CCSD and CCSD(T)\cite{Mazziotti:PRL08, Mazziotti:PRA10, Mazziotti:JCP12}.  Applications have been made to studying 
the energy barrier of between oxywater and hydrogen peroxide\cite{Mazziotti:JCP11b}, the relative populations of cis and trans 
carbonic acid at 210 K\cite{Mazziotti:JPCA11}, the diradical barrier to rotation of cis and trans 
diazene\cite{Mazziotti:JCP12b}, and the relative populations of olympicene and its isomers\cite{Mazziotti:JCP13}.  The 
algorithm was also explicitly 
constructed to preserve size extensivity, an important feature of any method intended for application to large systems.  
However, the asymptotic scaling of the algorithm is O(r$^6$), even when the excitation tensor is spectrally composed, due to 
the need for a linear number of terms in the spectral composition.  

In this paper, we show that the pRDM algorithm and the THC methodology can be combined to produce a tensor hypercontracted 
parametric reduced density matrix (THC-pRDM) algorithm that scales as O(r$^4$).  As in the work of Martinez et al., we will 
decompose both the ERI and the excitation tensors using THC.  But we will apply these decompositions to the pRDM algorithm 
of 
Mazziotti et al., rather than to more traditional electronic structure methods like CISD or CCSD.  We believe that our 
algorithm is the first example of a fully-implemented $O(r^4$) methodology with CCSD-like accuracy.  Our approach also has 
several intrinsic advantages.  First, because it produces the ground state 2-RDM, calculations of one- and two-particle 
properties of the ground state are straightforward, in contrast to the complexity of such calculations in coupled-cluster 
algorithms\cite{Monkhorst:IJQC77}.  Second, because RDM methods approach the problem of electronic structure differently 
than 
wavefunction-based 
methods, they provide a complementary perspective that can be valid when wavefunction-based methods are not.  Finally, the RDM 
approach suggests several approximations which are not available or at least are not obvious in wavefunction-based methods.  
We discuss a few of these in \refsec{Conclusions}.

Our paper is organized as follows: in \refsec{Theory}, we outline the theoretical underpinnings of our algorithm.  We review 
tensor hypercontraction and the pRDM algorithm and discuss several approximations that we make in the derivation of 
the THC-pRDM 
algorithm.  In \refsec{Results}, we show the results of applying our algorithm to several systems such as small molecules, 
hydrogen chains, and alkanes.  We find that our algorithm shows the expected behavior, approaching the standard pRDM result
as the number of auxiliary functions is increased.  We also show that we do indeed obtain 
$O(r^4)$ scaling for large systems, albeit 
with a large prefactor.  In \refsec{Conclusions}, we present some conclusions and discuss future directions for increasing the 
efficiency and accuracy of our method.

\section{Theory} \label{Section::Theory}
\subsection{THC decomposition of the Hamiltonian} \label{Section::THCERI}
The electronic Hamiltonian of an atom or molecule can be written in 2nd-quantized notation as
\begin{eqnarray} 
\hat{H} &=& \hat{H}_1 + \hat{H}_2 \\
\label{Equation::HDef}
&=&\sum_{ik}{\RDM{1}{\epsilon}^i_k \hat{c}^\dagger_i \hat{c}_k} + \sum_{ijkl}{\RDM{2}{\epsilon}^{ij}_{kl} \hat{c}^\dagger_i 
\hat{c}^\dagger_j \hat{c}_l \hat{c}_k }
\end{eqnarray}
where $h^i_k$ is the rank-2 matrix of 1-electron integrals and $\RDM{2}{\epsilon}^{ij}_{kl}$ is the rank-4 electronic 
repulsion integral tensor defined by
\begin{equation}
\RDM{2}{\epsilon}^{ij}_{kl} = \int{\int{d\v{r}_1 d\v{r}_2 \phi_i(\v{r}_1) \phi_k(\v{r}_1) 
\frac{1}{\abs{r_1-r_2}}\phi_j(\v{r}_2)\phi_l(\v{r}_2)}}
\end{equation}
The ERI tensor can be decomposed using a set of $P_H$ auxiliary functions as
\begin{eqnarray} 
\RDM{2}{\epsilon}^{ij}_{kl} &=& (ik|jl) \\
\label{Equation::HTHCDef}
&=& \sum_{P,Q=1}^{P_H}{h_{iP}h_{kP}J_{PQ}h_{jQ}h_{lQ}}.
\end{eqnarray}
Martinez and coworkers have developed a number of approaches to find an efficient THC decomposition of the ERI 
tensor\cite{Martinez:JCP12a,Martinez:JCP12b,Martinez:ARX13}.  We chose to 
implement this decomposition in terms of a simple non-linear least-squares fitting procedure.
First, we used the standard RI-V method to express the ERI tensor in terms of two rank-3 tensors,
\begin{equation} \label{Equation::RIV}
\RDM{2}{\epsilon}^{ij}_{kl} = \sum_{\mu,\nu=1}^{P_{RI}}{v^{ik}_\mu w_{\mu\nu} v^{jl}_\nu},
\end{equation}
where $\mu$ and $\nu$ label the $P_{RI}$ auxiliary density functions using the RI auxiliary basis found in the EMSL 
basis set exchange\cite{EMSL}. In the second step, we find some optimal set of auxiliary functions to decompose the 
rank-three tensor $v^{ik}_\mu$,  
\begin{equation}
v^{ik}_\mu = \sum_{P=1}^{P_H}{h_{iP}h_{kP}u_{\mu P}}.
\end{equation}
This decomposition can be accomplished through optimization of the cost-function
\begin{equation} \label{Equation::JRITHC}
J = \sum_{ik\mu}{\left( v^{ik}_\mu - \sum_{P=1}^{P_H}{h_{iP}h_{kP}u_{\mu P}}\right)^2}
\end{equation}
with respect to the variables $h_{iP}$ and $u_{\mu P}$.  The evaluation of the cost function and its gradients requires 
$O(r^2 P_{RI} P_{H})$ operations, although optimization can require several thousand iterations.  Having found an optimal 
$h_{iP}$ and $u_{\mu P}$ using a non-linear optimization scheme, the final THC 
decomposition is given by
\refeqn{HTHCDef} where
\begin{equation}
J_{PQ} = \sum_{\mu,\nu=1}^{P_{RI}}{u_{\mu P}u_{\nu Q} w_{\mu\nu}}
\end{equation}
Provided that $P_{RI}$ and $P_{H}$ both scale as $O(r)$, this entire process requires $O(r^4)$ operations, which has the same  
complexity as our THC-pRDM algorithm.  In practice, we find that if $P_{RI}$ is between $2r$ and $4r$ and $P_{H}$ is between 
$4r$ and $6r$, the THC decomposition leads to errors less than $2\u{mH}$ in the ground state energies obtained, which is 
suffucient for our purposes.  Table I 
shows a few representative molecules from our example and the ground state correlation energies obtained using the 
THC-compressed ERIs versus the exact ERIs.  

\begin{table} 
\begin{tabular}{|l||c|c|c|c|} 
\hline
Molecule & $P_H$(THC) & $P_H$(exact) & $E_c$(THC) & $E_c$(exact) \\
\hline 
CO(cc-pVDZ,$P_A=50$) & 194 & 465 & 308.17 & 307.81 \\
\hline 
H2O(cc-pVTZ, $P_A=100$) & 292 & 2192 & 286.44 & 285.19 \\
\hline 
CH4(cc-pVDZ, $P_A=50$) &  210 & 630 & 190.67 & 190.62 \\
\hline 
H8(cc-pVDZ, $P_A=48$) & 178 & 820 & 171.34 & 171.18 \\
\hline 
\end{tabular} 
\caption{Number of auxiliary functions $P_H$ and correlation energies $E_c$ (in $\u{mH}$) using the THC compressed ERI 
tensor versus the exact ERI tensor.  In all four cases, the 
error of the THC approximation was less than $2\u{mH}$.
}
\end{table}

\subsection{THC decomposition of the excitation tensor}
Configuration interaction and coupled cluster methods both rely on the use of an excitation operator which acts on some 
single-configuration Hartree-Fock reference wavefunction.  If we let $i,j,k,l$ label occupied orbitals and $a,b,c,d$ label 
unoccupied orbitals in the reference state, then the excitation operator can be written as
\begin{eqnarray} \label{TDef}
\hat{T}&=&\hat{T}_1+\hat{T}_2 \\
&=&\sum_{ia}{\RDM{1}{T}^a_i \hat{c}^\dagger_a \hat{c}_i} + \sum_{ijab}{\RDM{2}{T}^{ab}_{ij} \hat{c}^\dagger_a 
\hat{c}^\dagger_b \hat{c}_j \hat{c}_i}.
\end{eqnarray}
Spin symmetry can be taken into account explicitly; however, in this equation and what follows, we will ignore spin in order 
to simplify notation.  For the THC-pRDM method, we will apply THC decomposition to the two-particle excitation operator 
$\hat{T}_2$, modifying it slightly to take into account the differences between occupied and virtual orbitals.  We then obtain
\begin{equation} \label{Equation::T2THC}
\RDM{2}{T}^{ab}_{ij} = \sum_{R,S=1}^{P_A}{\left(y_{aR}x_{iR}z_{RS}y_{bS}x_{jS}-y_{aR}x_{jR}z_{RS}y_{bS}x_{iS}\right)}.
\end{equation}
Note that the second term in \refeqn{T2THC} is included to ensure the antisymmetry of the tensor $\hat{T}_2$.  The 
symmetry requirements of $\hat{T}_2$ also imply that $z_{RS}$ should be a symmetric matrix.  By writing the excitation tensor 
in terms of a THC decomposition, we have been able to compress a rank-4 object $\RDM{2}{T}^{ab}_{ij}$ into the 
product of three rank-2 objects, $x_{iR}, 
y_{aR},$ and $z_{RS}$.  A key to reducing the overall scaling of the electronic structure algorithm will be to determine how many auxiliary functions $P_A$ are 
needed to approximate the $\RDM{2}{T}$ tensor 
with sufficient accuracy.  In \refsec{Results}, we will show that we need only a linear number of auxiliary functions $P_A=O(r)$ to capture essentially 
all of the electronic correlation recovered by the standard pRDM algorithm, even 
though an exact representation of $\RDM{2}{T}$ would require $P_A = P(r^2)$ auxiliary functions.

\subsection{The pRDM algorithm} \label{Section::pRDM}
The version of the pRDM algorithm that we will employ here was first introduced in \cite{Mazziotti:PRL08} and subsequently 
refined in 
\cite{Mazziotti:PRA10}.  The algorithm is based on the construction 
of an optimal ground state 2-particle reduced density matrix (2-RDM), constrained to have a particular form parametrized by 
tensors $\RDM{1}{T}$ and $\RDM{2}{T}$.  The 
easiest way to understand the pRDM algorithm is to recognize that the 1- and 2-RDMs corresponding to a CISD wavefunction can 
be written exactly in terms of the 
excitation tensors $\RDM{1}{T}$ and $\RDM{2}{T}$.  

Again using the notation that indices $i,j,k,l$ represent occupied orbitals and $a,b,c,d$ represent unoccupied orbitals, the RDMs for the Hartree-Fock reference 
state are
\begin{eqnarray}
\RDM{1}{D}^{i}_{k} &=& \delta^i_k \\
\RDM{2}{D}^{ij}_{kl} &=& \delta^i_k\delta^j_l - \delta^i_l\delta^j_k
\end{eqnarray}
with all other elements equal to $0$.  Excitation out of this reference state adds cumulant components to the RDMs, which 
can be expressed in terms of $\RDM{1}{T}$ and $\RDM{2}{T}$. Straightforward 
evaluation from the CISD wavefunction yields

\begin{eqnarray} 
\label{Equation::RDMCISD1}
\RDM{1}{D}^{i}_{j} &=& \delta^i_j+\RDM{1}{\Delta}^i_j \\
\label{Equation::RDMCISD2}
\RDM{1}{D}^{a}_{i} &=& \RDM{1}{\Delta}^a_i \\
\label{Equation::RDMCISD3}
\RDM{1}{D}^{a}_{b} &=& \RDM{1}{\Delta}^a_b \\
\label{Equation::RDMCISD4}
\RDM{2}{D}^{ij}_{kl} &=& \delta^i_k\delta^j_l-\delta^i_l\delta^j_k+
\RDM{1}{\Delta}^i_k\delta^j_l-\RDM{1}{\Delta}^i_l\delta^j_k-\RDM{1}{\Delta}^j_k\delta^i_l+\RDM{1}{\Delta}^j_l\delta^i_k
+\RDM{2}{\Delta}^{ij}_{kl}  \\
\label{Equation::RDMCISD5}
\RDM{2}{D}^{ia}_{jk} &=& \delta^i_j\RDM{1}{\Delta}^a_k-\delta^i_k\RDM{1}{\Delta}^a_j+\RDM{2}{\Delta}^{ia}_{jk} \\
\label{Equation::RDMCISD6}
\RDM{2}{D}^{ia}_{jb} &=& \delta^i_j\RDM{1}{\Delta}^a_b-\RDM{1}{T}^i_b\RDM{1}{T}^j_a+\RDM{2}{\Delta}^{ia}_{jb} \\
\label{Equation::RDMCISD7}
\RDM{2}{D}^{ab}_{kl} &=& \RDM{2}{\Delta}^{ab}_{kl}\\
\label{Equation::RDMCISD8}
\RDM{2}{D}^{ia}_{bc} &=& \RDM{2}{\Delta}^{ia}_{bc} \\
\label{Equation::RDMCISD9}
\RDM{2}{D}^{ab}_{cd} &=& \RDM{2}{\Delta}^{ab}_{cd}
\end{eqnarray}
where the cumulants are given by
\begin{eqnarray} 
\label{Equation::RDMCISC1}
\RDM{1}{\Delta}^{i}_{j} &=& -\frac{1}{2}\sum_{kab}{\RDM{2}{T}^{ab}_{jk}\RDM{2}{T}^{ab}_{ik}}-\sum_a{\RDM{1}{T}^a_i 
\RDM{1}{T}^a_j} \\
\label{Equation::RDMCISC2}
\RDM{1}{\Delta}^{a}_{i} &=& \sum_{bj}{\RDM{2}{T}^{ab}_{ij} \RDM{1}{T}^b_j}+\RDM{1}{T}^a_i \\
\label{Equation::RDMCISC3}
\RDM{1}{\Delta}^{a}_{b} &=& \frac{1}{2}\sum_{cij}{\RDM{2}{T}^{ac}_{ij}\RDM{2}{T}^{bc}_{ij}} +\sum_i{\RDM{1}{T}^a_i 
\RDM{1}{T}^b_i}\\
\label{Equation::RDMCISC4}
\RDM{2}{\Delta}^{ij}_{kl} &=& \frac{1}{2}\sum_{ab}{\RDM{2}{T}^{ab}_{ij}\RDM{2}{T}^{ab}_{kl}}  \\
\label{Equation::RDMCISC5}
\RDM{2}{\Delta}^{ia}_{jk} &=& \sum_b{\RDM{2}{T}^{ab}_{jk}\RDM{1}{T}^b_i}\\
\label{Equation::RDMCISC6}
\RDM{2}{\Delta}^{ia}_{jb} &=& -\sum_{kc}{\RDM{2}{T}^{ac}_{jk}\RDM{2}{T}^{bc}_{ik} }\\
\label{Equation::RDMCISC7}
\RDM{2}{\Delta}^{ab}_{ij} &=& \RDM{2}{T}^{ab}_{ij} \\
\label{Equation::RDMCISC8}
\RDM{2}{\Delta}^{ia}_{bc} &=& \sum_j{\RDM{2}{T}^{bc}_{ij}\RDM{1}{T}^a_j }\\
\label{Equation::RDMCISC9}
\RDM{2}{\Delta}^{ab}_{cd} &=& \frac{1}{2}\sum_{ij}{\RDM{2}{T}^{ab}_{ij}\RDM{2}{T}^{cd}_{ij}}
\end{eqnarray}

The total energy of the system can then be calculated by tracing the 1- and 2-RDMs of the system over the 1- and 2-particle components of the Hamiltonian in 
\refeqn{HDef},
\begin{equation} \label{Equation::ERDM}
E = \trace{\RDM{1}{D}\cdot\RDM{1}{\epsilon}} + \trace{\RDM{2}{D}\cdot\RDM{2}{\epsilon}}
\end{equation}

Although \refeqto{RDMCISD1}{RDMCISD9} are exact for CISD wavefunctions, modifying the cumulants in \refeqn{RDMCISC2} and 
\refeqn{RDMCISC7} while leaving the other cumulant unchanged can produce 
approximations to the ground state energy that are far more accurate than CISD\cite{Mazziotti:PRL08,Mazziotti:PRA10}.  
Using Cauchy-Schartz inequalities based on $N$-representability conditions\cite{Mazziotti:PRL08,Kollmar:JCP06}, we 
can 
generalize these two equations to include a functional of $\RDM{1}{T}$ and $\RDM{2}{T}$,
\begin{eqnarray} 
\label{Equation::RDM1Func}
\RDM{1}{\Delta}^{a}_{i} &=& \sum_{bj}{\RDM{2}{T}^{ab}_{ij} \RDM{1}{T}^b_j}+\RDM{1}{T}^a_i\sqrt{1-\sum_{p,q=0}^1{\RDM{3}{c}^p_q \,
\RDM{\{p,q\}}{\Sigma}^a_i}-\sum_{p,q=0}^1{\RDM{2}{c}^p_q\, \RDM{\{p,q\}}{\Sigma}^a_i}} \\
\label{Equation::RDM2Func}
\RDM{2}{\Delta}^{ab}_{ij} &=& \RDM{2}{T}^{ab}_{ij}\sqrt{1-\sum_{p,q=0}^2{\RDM{4}{c}^p_q\, \RDM{\{p,q\}}{\Sigma}^{ab}_{ij}} - \sum_{p,q=0}^1{\RDM{3}{c}^p_q\, 
\RDM{\{p,q\}}{\Sigma}^{ab}_{ij}}},
\end{eqnarray}
where the $\Sigma$ tensors can be written in terms of $\RDM{1}{T}$ and $\RDM{2}{T},$\cite{Mazziotti:PRA10}
\begin{eqnarray}
\RDM{\{0,0\}}{\Sigma}^{a}_{i} &=& \sum_{kc}{(\RDM{1}{T}^{c}_{k})^2} \\
\RDM{\{1,0\}}{\Sigma}^{a}_{i} &=& \sum_{c}{(\RDM{1}{T}^c_i)^2} \\
\RDM{\{0,1\}}{\Sigma}^{a}_{i} &=& \sum_{k}{(\RDM{1}{T}^a_k)^2} \\
\RDM{\{1,1\}}{\Sigma}^{a}_{i} &=& (\RDM{1}{T}^a_i)^2 \\
\RDM{\{0,0\}}{\Sigma}^{ab}_{ij} &=& \frac{1}{4}\sum_{klcd}{(\RDM{2}{T}^{cd}_{kl})^2} \\
\RDM{\{1,0\}}{\Sigma}^{ab}_{ij} &=& -\RDM{1}{\Delta}^i_i-\RDM{1}{\Delta}^j_j \\
\RDM{\{0,1\}}{\Sigma}^{ab}_{ij} &=& \RDM{1}{\Delta}^a_a+\RDM{1}{\Delta^b_b} \\
\RDM{\{2,0\}}{\Sigma}^{ab}_{ij} &=& \RDM{2}{\Delta}^{ij}_{ij} \\
\RDM{\{0,2\}}{\Sigma}^{ab}_{ij} &=& \RDM{2}{\Delta}^{ab}_{ab} \\
\RDM{\{1,1\}}{\Sigma}^{ab}_{ij} &=& -\RDM{2}{\Delta}^{jb}_{jb}-\RDM{2}{\Delta}^{ja}_{ja} 
-\RDM{2}{\Delta}^{ib}_{ib} -\RDM{2}{\Delta}^{ia}_{ia} \\
\RDM{\{2,1\}}{\Sigma}^{ab}_{ij} &=& \sum_c{(\RDM{2}{T}^{ij}_{ac})^2 + (\RDM{2}{T}^{ij}_{bc})^2} \\
\RDM{\{1,2\}}{\Sigma}^{ab}_{ij} &=& \sum_k{(\RDM{2}{T}^{ik}_{ab})^2 + (\RDM{2}{T}^{jk}_{ab})^2} \\
\RDM{\{2,2\}}{\Sigma}^{ab}_{ij} &=& (\RDM{2}{T}^{ab}_{ij})^2 
\end{eqnarray}
The square-root factor in \refeqs{RDM1Func}{RDM2Func} acts like a normalization constant for the elements of the 1- and 
2-RDMs.  Physical explanations of these quantities can be found in \cite{Mazziotti:PRA10}.  What is important to note is 
that the constants $c^{\{p,q\}}$ can be chosen arbitrarily to define different functionals that relate the excitation tensors 
$\RDM{1}{T}$ and $\RDM{2}{T}$ to the components of the 1- and 2-RDM.  For instance, the choice $\RDM{4}{c}^{\{0,0\}}=\RDM{3}{c}^{\{0,0\}}=\RDM{2}{c}^{\{0,0\}}=1$ with all other coefficients equal to 
zero yields precisely the CISD wavefunction\cite{Mazziotti:PRA10}.  Mazziotti and coworkers devised several other 
functionals, all of which are obtained by choosing different coefficients for $\RDM{4}{c}, \RDM{3}{c},$ and 
$\RDM{2}{c}$ (see Table II).  To perform an energy calculation, we select a functional and then minimize the energy in 
\refeqn{ERDM} with respect to the tensors 
$\RDM{1}{T}$ and $\RDM{2}{T}$.  In this paper, we 
will focus on the $M_2$ functional introduced by Mazziotti in \cite{Mazziotti:PRA10}, which was able to provide ground state energies with an accuracy somewhere between 
CCSD and CCSD(T) for a wide variety of molecular systems.  

\begin{table} 
\begin{tabular}{|l||c|c|c|c|c|c|c|c|c|c|c|c|} 
\hline 
Functional & $\RDM{4}{c}^0_0$ & $\RDM{4}{c}^1_0$ & $\RDM{4}{c}^2_0$ & $\RDM{4}{c}^1_1$ & $\RDM{4}{c}^2_1$ & $\RDM{4}{c}^2_2$ & 
$\RDM{3}{c}^0_0$ & $\RDM{3}{c}^1_0$ & $\RDM{3}{c}^1_1$ & $\RDM{2}{c}^0_0$ & $\RDM{2}{c}^1_0$ & $\RDM{2}{c}^1_1$ \\ 
\hline 
CISD & 1 & 0 & 0 & 0 & 0 & 0 & 1 & 0 & 0 & 1 & 0 & 0 \\ 
\hline 
CEPA & 0 & 0 & 0 & 0 & 0 & 0 & 0 & 0 & 0 & 0 & 0 & 0 \\ 
\hline 
$M_2$ & 0 & 1/5 & 3/5 & -1/10 & 0 & -3/5 & 0 & 1 & -1 & 0 & 1 & -1 \\ 
\hline 
$M'_2$ & 0 & 1/5 & 3/5 & -1/10 & 0 & 0 & 0 & 1 & -1 & 0 & 1 & -1 \\ 
\hline 
\end{tabular} 
\caption{Selected 2-RDM functionals defined in \cite{Mazziotti:PRA10} and in the current work.  Coefficients not shown are 
specified by particle/hole duality ($c^p_q = c^q_p$).  To preserve $O(r^4)$ computational scaling, a 
single term from the $M_2$ functional must be truncated, yielding the $M'_2$ functional.  Table III shows that this change 
produces negligible differences in energy.
}
\end{table}

One other distinction that will be important for the THC-pRDM algorithm is that the zeroth order and cumulant contributions to 
the total energy in \refeqn{ERDM} can be evaluated separately.  For this reason, we choose to evaluate the Hartree-Fock energy 
and density matrix using the exact 2-electron ERIs rather than the RI-V or THC approximations.  We invoke the THC 
approximation to the ERIs \emph{only} when evaluating the energetic contribution of the cumulants.  This choice is important 
since the absolute Hartree-Fock electronic energy is much larger than the correlation energy.  Because we restrict our use of 
the THC approximation to the calculation of the correlation energy, we need a far less precise approximation of the ERIs to 
obtain an approximation of the total energy with sub-miliHartree accuracy.

\subsection{Applying THC to the pRDM algorithm}
The application of tensor hypercontraction to the pRDM algorithm is relatively straightforward.  We first write the 
rank-4 ERI tensor as a THC decomposition, according to the methodology discussed in \refsec{THCERI}.  If we also express the 
rank-4 excitation tensor $\RDM{2}{T}$ as a THC decomposition (see \refeqn{T2THC}), then we can calculate the energy 
contributions of most of the cumulants in \refeqto{RDMCISC1}{RDMCISC9} in $O(r^4)$ operations.  For instance, the energy 
contribution from the cumulant in \refeqn{RDMCISC8} is
\begin{eqnarray}
E &=& \sum_{iabc}{\RDM{2}{\epsilon}^{ia}_{bc} \RDM{2}{\Delta}^{ia}_{bc}} \\ 
 &=& \sum_{PQRS}{\left(x'_{RP}T'_{SQ}y'_{RP}y'_{SQ}J_{PQ}z_{RS}-x'_{SP}T'_{RQ}y'_{RP}y'_{SQ}J_{PQ}z_{RS}\right)}
\end{eqnarray}
where
\begin{eqnarray}
x'_{RP} &=& \sum_{i}{x_{iR}h_{iP}} \\
y'_{RP} &=& \sum_{a}{y_{aR}h_{aP}} \\
T'_{RP} &=& \sum_{ia}{x_{iR}\RDM{1}{T}_i^a h_{aP}} .
\end{eqnarray}
The derivative of the energy expression with respect to the elements $x_{iR},y_{aR}$ and $z_{RS}$ of the excitation operator can likewise be calculated in 
$O(r^4)$ operations.  Only the cumulant 
contribution in \refeqn{RDM2Func} is potentially problematic and necessitates two simple approximations to preserve $O(r^4)$ 
scaling.  

First, we observe that the square-root operation in \refeqn{RDM2Func} is problematic because it inextricably entangles all 
four indices $a,b,i,j$ in the tensor $\RDM{2}{T}^{ab}_{ij}$ with the tensors $\RDM{\{p,q\}}{\Sigma}^{ab}_{ij}$. Unfortunately, 
the speed-up produced by the THC decomposition necessitates that these indices be disentangled into pairs of 
at most two indices.  To solve this problem, we 
first approximate the square-root operation via Taylor expansion\cite{Mazziotti:PRA10}.  Provided that the excitation tensor 
$\RDM{2}{T}$ is
relatively small in magnitude, which is the case for all systems 
close to their mean-field reference, 
this approximation should be negligible.  Next, we observe that the $\{2,1\}, \{1,2\}$ and $\{2,2\}$ terms in 
\refeqn{RDM2Func} entangle $3$, $3$, and $4$ indices, again eliminating the possibility of 
an $O(r^4)$ evaluation.  However, the $M_2$ functional fortuitously sets the coefficients $\RDM{4}{c}^{\{2,1\}}$ and $\RDM{4}{c}^{\{1,2\}}$ to zero.  The only 
approximation we need to make is to additionally set $\RDM{4}{c}^{\{2,2\}} = 0$ (see Table II).  This is not a significant 
approximation either, since all the 
elements of 
$\RDM{\{2,2\}}{\Sigma^{ab}_{kl}}$ should be small if the system is close to its 
mean-field reference because each element in the class contains only one positive term.  Table III shows that the effect of 
both of these approximations on the ground state energies for 
several representative systems is negligible, amounting to less than $0.2\u{mH}$.

\begin{table}
\begin{tabular}{|l||c|c|c|c|}
\hline
 & standard & Taylor & $M'_2$ & Taylor + $M'_2$ \\
\hline 
HCN(cc-pVDZ) & 313.05 & 313.08 & 312.91 & 312.95 \\
\hline
H2O(cc-pVTZ) & 288.43 & 288.44 & 288.41 & 288.42 \\
\hline
CH4(cc-pVDZ) & 192.39 & 192.40 & 192.38 & 192.39 \\
\hline
H8(cc-pVDZ) & 172.32 & 172.34 & 172.27 & 172.28  \\
\hline
\end{tabular}
\caption{Comparison of the pRDM ground state energy under various approximations.  The first column shows the result using the standard pRDM algorithm with the 
$M_2$ functional.  The second column uses a Taylor expansion of the functional $M_2$.  The third column uses the $M'_2$ functional with no Taylor expansion.  
The last column employs both approximations, corresponding to the approach taken by the THC-pRDM algorithm.  The 
approximations produce negigible differences in 
the final energy for the 
systems studied.}
\end{table}

\subsection{The initial guess for the excitation tensor} \label{Section::THCMP2}
One factor that can greatly improve algorithmic convergence is a choice of a good initial guess for the parameters 
$x_{iR},y_{aR}$ and $z_{RS}$ that compose the 
THC-pRDM excitation operator.  The MP2 amplitudes for a given 
Hamiltonian 
are known analytically to be
\begin{eqnarray}
\RDM{2}{T}^{ab}_{ij} &=& \frac{1}{F_{ii}+F_{jj}-F_{aa}-F_{bb}}\RDM{2}{\epsilon}^{ab}_{ij} \\
&=& \frac{1}{F_{ii}+F_{jj}-F_{aa}-F_{bb}}\sum_{P,Q=1}^{P_H}{h_{aP}h_{iP}J_{PQ}h_{bQ}h_{jQ}}
\end{eqnarray}
where $F$ is the Fock matrix.  However, we need to express these amplitudes in THC form. Here we 
follow \cite{Martinez:JCP12a} in 
writing the MP2 amplitudes as an integral and then applying Gauss-Laguerre quadrature to express them as a weighted sum over 
knot points.  This procedure yields
\begin{eqnarray}
\RDM{2}{T}^{ab}_{ij} &=& \frac{1}{F_{ii}+F_{jj}-F_{aa}-F_{bb}}\sum_{P,Q=1}^{P_H}{h_{aP}h_{iP}J_{PQ}h_{bQ}h_{jQ}} \\ 
&=& \int_0^\infty{dx \sum_{P,Q=1}^{P_H}{h_{aP}h_{iP}J_{PQ}h_{bQ}h_{jQ}}\exp{\left(-(F_{ii}+F_{jj}-F_{aa}-F_{bb})x\right)}} \\ 
&=& \frac{1}{2\Delta E}\int_0^\infty{dx' \sum_{P,Q=1}^{P_H}{h_{aP}h_{iP}J_{PQ}h_{bQ}h_{jQ}}\exp{\left(-(F_{ii}+F_{jj}-F_{aa}-F_{bb}-2\Delta E)x'/2\Delta E 
\right)\exp{(-x')} }} 
\\ 
&\approx& \frac{1}{2\Delta E}\sum_{n=1}^{n_k}{w_n \sum_{P,Q=1}^{P_H}{h'_{aP,n}h'_{iP,n}J_{PQ}h'_{bQ,n}h'_{jQ,n} } } 
\end{eqnarray}
where
\begin{eqnarray}
h'_{aP,n} &=& \exp{\left((F_{aa}-E_{LUMO})x'_n/2\Delta E\right) h_{aP}} \\
h'_{iP,n} &=& \exp{\left(-(F_{ii}-E_{HOMO})x'_n/2\Delta E\right) h_{iP}} \\
\Delta E &=& E_{HOMO}-E_{LUMO}
\end{eqnarray}
and where $(x'_n,w_n)$ are a set of $n_k$ Gauss-Laguerre knot points and weights.  From this expression, we can then define a cost function
\begin{eqnarray}
J &=& \sum_{abij}{\left(\RDM{2}{T}^{ab}_{ij}-\sum_{R,S=1}^{P_A}{y_{aR}x_{iR}z_{RS}y_{bS}x_{jS}}\right)^2} \\
\label{Equation::JMP2}
&\approx& \sum_{abij}{\left(\frac{1}{2\Delta E}\sum_{n=1}^{n_k}{w_n \sum_{P,Q=1}^{P_H}{h'_{aP,n}h'_{iP,n}J_{PQ}h'_{bQ,n}h'_{jQ,n}}}-
\sum_{R,S=1}^{P_A}{y_{aR}x_{iR}z_{RS}y_{bS}x_{jS}}\right)^2}.
\end{eqnarray}
\refeqn{JMP2} and the derivatives of $J$ with respect to $x_{iR},y_{aR},$ and $z_{RS}$ can be evaluated in $O(n_k^2 P_A P_H^2)$ operations.  We find that $n_k = 
8$ 
provides sufficient accuracy for our purposes.  And since $P_A$ and $P_H$ both scale as $r$, the entire 
optimization procedure will scale as $O(r^3)$.  

Because the MP2 amplitudes are not normalized, our initial condition must include a normalization factor
$z_{RS} \to \alpha z_{RS}$ that prevents the excitation amplitudes from growing increasingly large as the system size 
increases.  We find that 
by using this guess, the initial MP2 guess will already recover a large fraction of the correlation energy and can 
dramatically reduce the convergence time of our algorithm.

\subsection{The final THC-pRDM algorithm}
Having discussed the various constituents of the THC-pRDM algorithm, we can now provide a complete outline:
\begin{enumerate}
\item Calculate the 1- and 2-electron integrals using a standard electronic structure package, such as QM4D, the package we 
used\cite{QM4D}.  Use RI-V to also 
approximately express the ERIs in 
terms of auxiliary density functions (see \refeqn{RIV})
\item Minimize the cost function in \refeqn{JRITHC} to obtain a good THC decomposition of the ERIs from the RI-V 
calculation
\item Perform standard Hartree-Fock using the exact ERIs to obtain the ground state energy and density matrix
\item Obtain the initial starting values for the matrices $x_{iR}$, $y_{aR}$, and $z_{RS}$ from the approximate 
MP2 amplitudes (see \refsec{THCMP2})
\item Select a normalization factor $\alpha$ which minimizes the energy of the initial MP2 guess
\item Minimize the energy with respect to $x_{iR}$, $y_{aR}$, $z_{RS}$, and $\RDM{1}{T}^a_i$ using a non-linear 
optimization algorithm.  Convergence can be improved by optimizing the variables in three steps: $(x_{iR},y_{aR})$, 
$z_{RS}$, and $\RDM{1}{T}^a_i$.  The optimization process is repeated until a sufficient level of convergence is achieved.
\end{enumerate}

\section{Results} \label{Section::Results}

We applied the THC-pRDM algorithm to a variety of molecular systems to test its accuracy.  Our first concern was how quickly 
the THC-pRDM algorithm would approach the standard pRDM limit as we increased the number of auxiliary functions $P_A$.  
Table 
IV shows the 
results 
of our calculations for a set of six small molecules in the cc-pVDZ 
basis.  As the number $P_A$ of auxiliary functions is increased, the correlation energy does rapidly approach that of the 
standard pRDM algorithm.  Table V shows the same six molecules in the cc-vPTZ basis and the same behavior is observed.  In 
this case, the largest number of auxiliary functions yields an answer with greater accuracy than CCSD, except for the HCN 
molecule.  It is encouraging to see that only about $P_A = 1.5 r$ auxiliary functions are needed to obtain accuracy similar to 
that of CCSD in both basis sets.   Still, the high symmetry of these small molecules and the fact that the number of electrons 
remains constant as the basis size is increased makes it difficult to extract any scaling information from these test cases.

\begin{table}
\begin{tabular}{|l||c|c|c|c|c|c|c|}
\hline
Molecule & $P_A=20$ & $P_A=30$ & $P_A=40$ & $P_A=50$ & $P_A=\infty$ & 
CCSD & CCSD(T) \\
\hline 
CH$_2$ & 139.91 & 143.31 & 144.19 & 144.62 & 145.09 & 142.38 & 145.45 \\
\hline
CO & 277.91 & 294.84 & 303.52 & 308.17 & 312.38 & 306.88 & 318.50 \\
\hline
H$_2$O & 207.42 & 215.28 & 216.94 & 217.87 & 218.65 & 217.33 & 220.62 \\
\hline
HCN & 274.69 & 293.30 & 300.39 & 307.08 & 313.05 & 307.18 & 320.20 \\
\hline
N$_2$ & 295.98 & 312.79 & 323.08 & 326.07 & 331.78 & 326.32 & 339.71 \\
\hline
NH$_3$ & 191.59 & 203.65 & 207.71 & 209.04 & 210.34 & 208.23 & 212.28 \\
\hline
\end{tabular}
\caption{Ground state correlation energies in mH for small molecules in the cc-pVDZ basis.  As the number of auxiliary functions is increased, the THC-pRDM 
energy approaches the standard pRDM 
energy, 
which lies between the CCSD and CCSD(T) energies.
}
\end{table}

\begin{table}
\begin{tabular}{|l||c|c|c|c|c|c|c|}
\hline
Molecule & $P_A=40$ & $P_A=60$ & $P_A=80$ & $P_A=100$ & $P_A=\infty$ & 
CCSD & CCSD(T) \\
\hline 
CH$_2$ & 173.31 & 180.13 & 181.90 & 182.41 & 183.27 & 179.36 & 184.45 \\
\hline
CO & 355.35 & 381.59 & 389.00 & 392.48 & 399.61 & 391.45 & 409.53 \\
\hline
H$_2$O & 272.69 & 281.02 & 284.77 & 286.44 & 288.40 & 285.40 & 293.40 \\
\hline
HCN & 349.78 & 370.50 & 377.44 & 384.94 & 395.95 & 387.00 & 406.15 \\
\hline
N$_2$ & 378.67 & 401.71 & 408.98 & 411.24 & 421.66 & 411.05 & 431.40 \\
\hline
NH$_3$ & 251.65 & 263.16 & 266.49 & 267.91 & 270.25 & 266.51 & 274.47 \\
\hline
\end{tabular}
\caption{Ground state correlation energies in mH for small molecules in the cc-pVTZ basis.  The number of basis functions $r$ is roughly twice as large as in 
Table 3 and the molecules require roughly twice as many auxiliary functions $P_A$ to achieve the same accuracy.
}
\end{table}

A more important test of scaling comes from the alkane series shown in Table VI.  In this table, we report the percentage of 
the CCSD(T) correlation energy that can be obtained from THC-pRDM, standard pRDM and CCSD.  In this table, we can 
clearly see that the number of auxiliary functions needed to obtain any particular level of accuracy scales linearly with $N$, 
the length of the carbon chain.  The number of auxiliary functions needed to obtain CCSD-level accuracy also represents a huge 
compression of the excitation tensor.  For instance, the excitation tensor for C$_4$H$_{10}$ contains approximately 3 
million 
independent 
parameters. In contrast, the THC compressed excitation operator with $P_A=200$ auxiliary functions contains approximately 
sixty thousand independent elements, yielding a compression factor of $50$.  Yet despite this compression, the THC-pRDM 
algorithm easily attains accuracy greater than the CCSD result.  

Similar behavior is shown in Table VII, where the algorithm was 
applied to linear hydrogen chains where the internuclear separation of each hydrogen was $R=0.74\u{\AA}$. Once again, the 
THC-pRDM algorithm required a number of auxiliary functions that scaled linearly with the number of atoms in the chain.  Both 
of these results indicate that the formal scaling of the THC-pRDM algorithm is indeed $O(r^4)$.  This scaling is clearly 
visible in Figure 1, where the computational cost of the hydrogen chain is plotted as a function of 
basis set size.  Despite the large prefactor, the system shows $O(r^4)$ scaling, as expected.

\begin{table}
\begin{tabular}{|l||c|c|c|c|c|c|c|c|c|}
\hline
Molecule & $N$ & $P_A=10 N$ & $P_A=20 N$ & $P_A=30 N$ & $P_A=40 N$ & $P_A=50 N$ & $P_A=\infty$ & CCSD & CCSD(T) \\
\hline 
CH$_4$ & 1 & 65.1\% & 88.4\% & 96.3\% & 97.9\% & 98.4\% & 99.3\% & 98.0\% & 100.0\% \\
\hline
C$_2$H$_6$ & 2 & 68.9\% & 90.2\% & 96.3\% & 97.8\% & 98.3\% & 99.1\% & 97.6\% & 100.0\% \\
\hline
C$_3$H$_8$ & 3 & 70.1\% & 90.7\% & 95.7\% & 97.2\% & 98.2\% & 99.0\% & 97.4\% & 100.0\% \\
\hline
C$_4$H$_{10}$ & 4 & 68.4\% & 89.9\% & 95.8\% & 97.6\%  & 98.1\% & 98.9\% & 97.3\% & 100.0\% \\
\hline
\end{tabular}
\caption{Percentage of the CCSD(T) correlation energy recovered by each method for an alkane molecule.  This table shows that the number of auxiliary functions 
needed to recover some 
constant percentage of the CCSD(T) correlation energy scales linearly with the size of the molecule.  More practically, in all cases $P_A = 40 N$ auxiliary 
functions are sufficient to achieve CCSD accuracy with the THC-pRDM algorithm.
}
\end{table}

\begin{table}
\begin{tabular}{|l||c|c|c|c|c|c|c|c|c|c|}
\hline
Molecule & $N$ & $P_A=N$ & $P_A=2 N$ & $P_A=3 N$ & $P_A=4 N$ & $P_A=5 N$ & $P_A=6 N$ & $P_A=\infty$ & CCSD & CCSD(T) \\
\hline 
H$_2$ & 2 & 69.0\% & 95.9\% & 98.0\% & 99.9\% & 100.0\% & 100.0\% & 100.0\% & 100.0\% & 100.0\% \\
\hline
H$_4$ & 4 & 61.6\% & 84.9\% & 96.2\% & 99.0\% & 99.6\% & 99.9\% & 100.2\% & 99.2\% & 100.0\% \\
\hline
H$_6$ & 6 & 57.6\% & 81.3\% & 95.1\% & 97.9\% & 99.6\% & 99.8\% & 100.3\% & 98.7\% & 100.0\% \\
\hline 
H$_8$ & 8 & 55.8\% & 78.5\% & 94.0\% & 97.8\% & 99.3\% & 99.8\% & 100.5\% & 98.4\% & 100.0\% \\
\hline 
H$_{10}$ & 10 & 58.2\% & 79.8\% & 93.8\% & 97.6\% & 99.0\% & 99.7\% & 100.4\% & 98.1\% & 100.0\% \\
\hline 
H$_{12}$ & 12 & 56.3\% & 80.2\% & 93.6\% & 97.3\% & 99.5\% & 100.1\% & 100.4\% & 97.9\% & 100.0\% \\
\hline 
H$_{14}$ & 14 & 57.7\% & 79.9\% & 92.8\% & 96.5\% & 98.9\% & 99.7\% & 100.4\% & 97.7\% & 100.0\% \\
\hline 
H$_{16}$ & 16 & 54.1\% & 78.8\% & 92.3\% & 96.2\% & 98.8\% & 99.4\% & 100.4\% & 97.6\% & 100.0\% \\ 
\hline
\end{tabular}
\caption{Percentage of the CCSD(T) correlation energy recovered by each method for a linear chain of hydrogen atoms. In all cases $P_A = 5N$ auxiliary functions are 
sufficient to recover as much correlation energy as CCSD.
}
\end{table}

\begin{figure}
\includegraphics[width=3in]{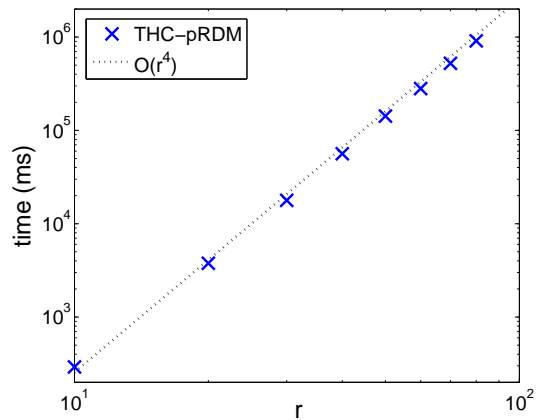}
\caption{The computational cost of a single energy and gradient evaluation for the linear H chains studied in 
\refsec{Results}.  The time in $\u{ms}$ is plotted as a function of the number of basis functions $r$.  
The cost has an $O(r^4)$ dependence, as expected from the linear scaling of $P_A$ and $P_H$.}
\end{figure}

\section{Conclusions} \label{Section::Conclusions}
In this article, we introduced the tensor-hypercontracted paramatric reduced density matrix method (THC-pRDM) for 
electronic structure calculation.  The method combines the tensor hypercontraction scheme of Martinez et 
al \cite{Martinez:JCP12a} with the parametric RDM method of Mazziotti et al \cite{Mazziotti:PRL08} to produce a formally 
$O(r^4)$ method that has an accuracy comparable to CCSD/CCSD(T).  We applied our method to several small molecules in the 
cc-pVDZ and cc-pVTZ basis sets, to alkanes, and to hydrogen chains.  In the latter two cases, it can be clearly seen that 
the 
number of auxiliary functions needed to provide a good approximation of the ERI tensor and the excitation tensor $\RDM{2}{T}$ 
scales 
linearly with the system size, guaranteeing an overall $O(r^4)$ scaling.

There are several steps in our algorithm which can be improved.  First, we accomplished the fitting of the ERI tensor to a
THC form in a manner that was straightforward but not necessarily optimal.  Martinez and coworkers have developed approaches 
that use a larger number of auxiliary functions, but obtain the fitting parameters $h_{iP}$ and $J_{PQ}$ 
more efficiently.  Given that we were able to use a small number of auxiliary functions, our hope is that some compromise 
can be struck between the size of the auxiliary basis and the speed with which the fitting can be performed.  Second, we used 
the MP2 excitation amplitudes to obtain an initial guess for our optimization algorithm.  Although this initial guess was far 
better than a random initial condition, there is again no guarantee that it is optimal.  Future work can be done to provide an 
initial condition that both converges rapidly and avoids becoming trapped in local minima.

The most important area for improvement has to do with the calculation of the energy and gradient given the 
THC parameters.  Figure 1 shows, despite the efficient scaling of the THC-pRDM algorithm, that there is a large prefactor 
associated with each evaluation of the enegy and gradient.  This large prefactor presently restricts the application 
of our method to systems similar in size to the ones studied, consisting of approximately $r=100$ basis functions.  Obviously, we 
would like to reduce this prefactor dramatically to increase the range of applicability of our algorithm.  

Apart from technical improvements to code writing and compilation, there are several physical approximations that might yield 
improved speed.  First, the inclusion of single excitations contributes to a fairly large fraction of the computational cost.  
Because the effect of singles excitations is usually much smaller than the effect of doubles excitations, we may be able to 
iognore the singles contribution entirely or to take it into account by modifying the functional.  Second, because each 
cumulant from \refeqto{RDMCISC1}{RDMCISC9} contributes to the energy independently, we can identify cumulants that either 
contribute little to the energy or are closely correlated with other cumulants.  In the future, we will investigate the 
possibility of approximating small but costly cumulants as functionals of less-expensive cumulants.  Finally, the THC form of 
the excitation operator is not necessarily ideal, especially for spatially extended systems.  To improve cost, we could 
further restrict the excitation operator by constraining excitations to only act ``locally."  This approach would reduce the 
size of the parameter space over which we optimize.  

In conclusion, we believe that the THC-pRDM algorithm presents a new and interesting approach to electronic structure 
calculations, one that presents numerous areas of new research and the potential of becoming an accurate and 
practically-relevant method for evaluation of the ground state properties of large systems.

\begin{acknowledgments}
NS and WY would like to acknowledge support from the UNC EFRC: Solar Fuels and Next 
Generation Photovoltaics, an Energy Frontier Research Center funded by 
the U.S. Department of Energy, Office of Science, Office of Basic Energy 
Sciences under Award Number DE-SC0001011 and the National Science Foundation (NSF)(CHE-09-11119).  D.A.M. gratefully 
acknowledges the National Science Foundation under Award Number CHE-1152425 and the Army Research Office under Award Number 
W91 INF-1 1-504 1-0085.  HvA thanks the FWO-Flanders for support.
\end{acknowledgments}


\end{document}